\documentclass[letterpaper,10pt]{article} 

\usepackage{osameet3} 

\usepackage{amsmath,amssymb}
\usepackage[colorlinks=true,bookmarks=false,citecolor=blue,urlcolor=blue]{hyperref} 

\usepackage{subfigure}
\usepackage{multicol}

\begin{document}
\title{Quality of Transmission Estimation for Network Planning: how to handle single-channel nonlinear effects?}
\vspace{-0.5cm}
\author{Andrea D'Amico$^{(1)}$, Elliot London$^{(1)}$, Emanuele Virgillito$^{(1)}$, Antonio Napoli$^{(2)}$, Vittorio Curri$^{(1)}$}
\address{$^{(1)}$DET - Politecnico di Torino, ITALY; $^{(2)}$Infinera, Sankt-Martin-Str. 76, 81541 Munich, Germany}
\email{andrea.damico@polito.it}
\copyrightyear{2020}
\vspace{-0.5cm}
\begin{abstract}
We propose an analytical method to evaluate the equivalent SPM component of the NLI generated by each fiber span, to enabling a fully disaggregated evaluation of the GSNR degradation, as required in network planning.
\end{abstract}
\vspace{-3mm}
\section{Introduction}
\vspace{-2mm}
\begin{figure*}[b!]
    \centering
    \includegraphics[width=0.75\textwidth]{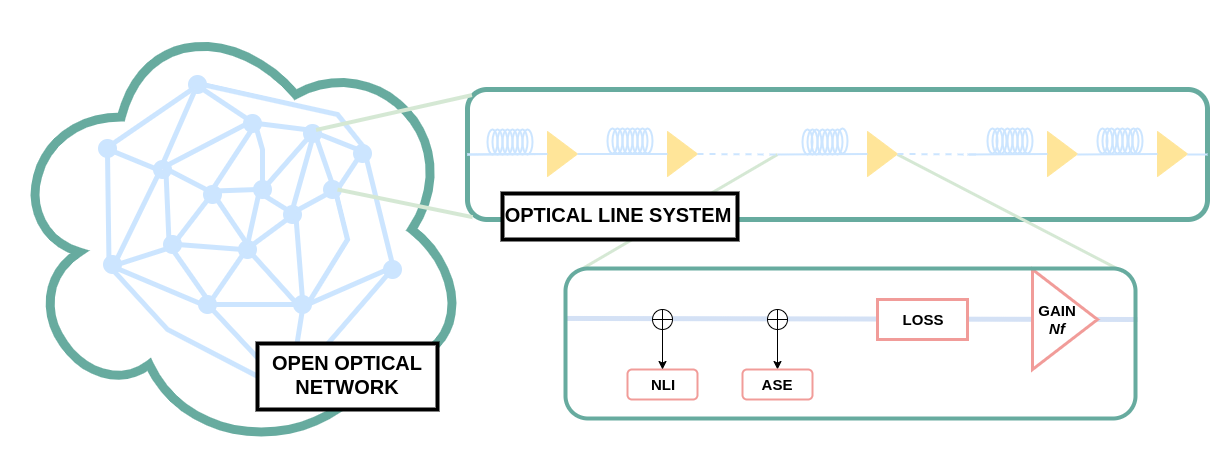}
    \vspace{-0.35cm}
   \caption{Disaggregated abstraction of an optical network for QoT purposes.}
   \label{Fig:disaggregated}
\end{figure*}
Optical networks are fast moving towards partial disaggregation, with a possible final evolution being full disaggregation.
As a result, optical line systems (OLS) from different vendors may be required to operate within the same infrastructure in order to transport physical lightpaths operated by multivendor transponders.
In this scenario, starting from planning, network management operations are enabled by establishing common API's and data structures down to the transport layer.
In this work, we investigate 
the abstraction of the transport layer for planning purposes as described in Fig.~\ref{Fig:disaggregated}, focusing on 
single channel nonlinear interference (NLI).

It is well accepted that the unique meter to predicting the quality-of-transmission (QoT) in state-of-the-art networks deploying coherent optical technologies is the generalized signal-to-noise ratio (GSNR), considering both the effects of the accumulated ASE noise and NLI~\cite{filer18multi}.
So, the physical layer of an optical network can be abstracted as a graph whose weights on edges and nodes are the GSNR degradation due to lightpaths' (LP) propagation on OLS's and at ROADM nodes, respectively~\cite{curri_jlt}.
Note that the GSNR corresponds to the error vector magnitude (EVM)~\cite{{evm}} on the DSP-recovered constellation in case of additive Gaussian Noise channel, as in the present work.
For network planning and controlling, the physical layer abstraction must be \emph{local}, and network elements responsible for disturbance generation -- fiber spans for the NLI and amplifiers for the ASE noise -- must be considered independently, with a fully disaggregated approach~\cite{auge2019open}, as illustrated in Fig.~\ref{Fig:disaggregated}.
Moreover, to avoid out-of-service in progressive LP deployment, the predicted GSNR degradation must be conservative for all the possible deployed traffic: Full loading is indeed a conservative choice.
Regarding disturbances, the ASE noise is indeed a \emph{local} effect depending on the characteristics of the deployed amplifiers.
The accumulation of the NLI can be separated into the independent contributions of cross-phase- (XPM) and self-phase-modulation (SPM)~\cite{Dar2013,egn,icton19}.
The XPM generation is well approximated as a span by span statistical incoherent effect for all realistic scenarios~\cite{ucl14,icton19}.
So, the XPM contribution is \emph{local} too, in agreement with the fully disaggregated abstraction of Fig.~\ref{Fig:disaggregated}.
In this scenario, the only obstacle to the \emph{local} abstraction is the SPM component of the NLI. For each fiber span, this is a random process statistically correlated to the SPM generated by the previous crossed fiber spans on the route followed by the LP under test~\cite{ptl12,ucl14}.
The overall amount of disturbance, in power $\mathrm{P}_{dist}^{(n)}$, generated on a single channel by the fiber span~$(n)$ within the LP route can be expressed as
\begin{equation}\label{Eq: nonlocality}
    \mathrm{P}_{dist}^{(n)}=\mathrm{P}_{ASE}^{(n)}+\mathrm{P}_{XPM}^{(n)}+\mathrm{P}_{SPM}^{(n)} +\sum_{k=1}^{n-1}C_{n,k}\mathrm{P}_{SPM}^{(k)} = \mathrm{P}_{ASE}^{(n)}+\mathrm{P}_{XPM}^{(n)}+\mathrm{P}_{SPM}^{(n)}+ \mathrm{COH}_{SPM}(n) \;,
\end{equation}
where $k$ is the index on the route spans and $C_{n,k}$ are the $k$-span to $n$-span cross-correlation coefficients, $\mathrm{P}_{SPM}^{(k)}$ is the power of the SPM noise generated at the $k$-span; both the latter quantities depend on lengths ($L_s$) and dispersion coefficients ($\beta_2$) of each crossed fiber span, as well as on the symbol rate ($R_s$). On the right hand side of Eq.\ref{Eq: nonlocality}, we enclose in $\mathrm{COH}_{SPM}(n)$ the entire coherent accumulation responsible for the \emph{nonlocality} of the SPM disturbance. 
In this work, we propose a mathematical method validated by accurate split-step simulations enabling a local \emph{equivalent} abstraction of the SPM by considering its asymptotic correlation.
We demonstrate that the approach is conservative for realistic scenarios on SMF and LEAF fibers
and properly scales with system parameters through the factor $\theta=\pi R_s^2\beta_2 L_s$.
Therefore, it enables a disaggregated, accurate yet conservative approach also for the SPM component of the NLI. This contribution will become more significant as the symbol rate increases, such as for 400G and 800G transceivers~\cite{cit400}.
\section{Theory and Validation}
\vspace{-2mm}
\begin{figure*}[b!]
    \centering
    \begin{subfigure}(a)
        \centering
        \includegraphics[width=0.45\textwidth]{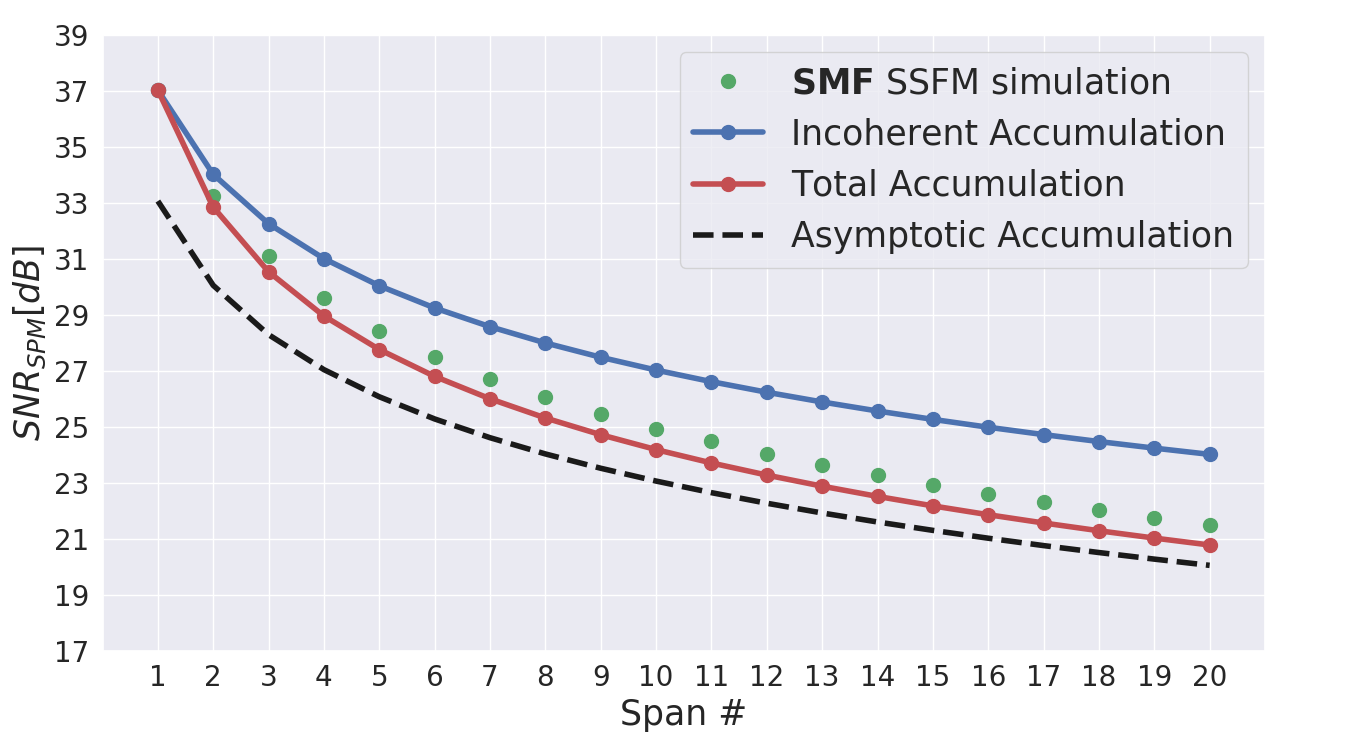}
    \end{subfigure}
    ~
    \begin{subfigure}(b)
        \centering
        \includegraphics[width=0.449\textwidth]{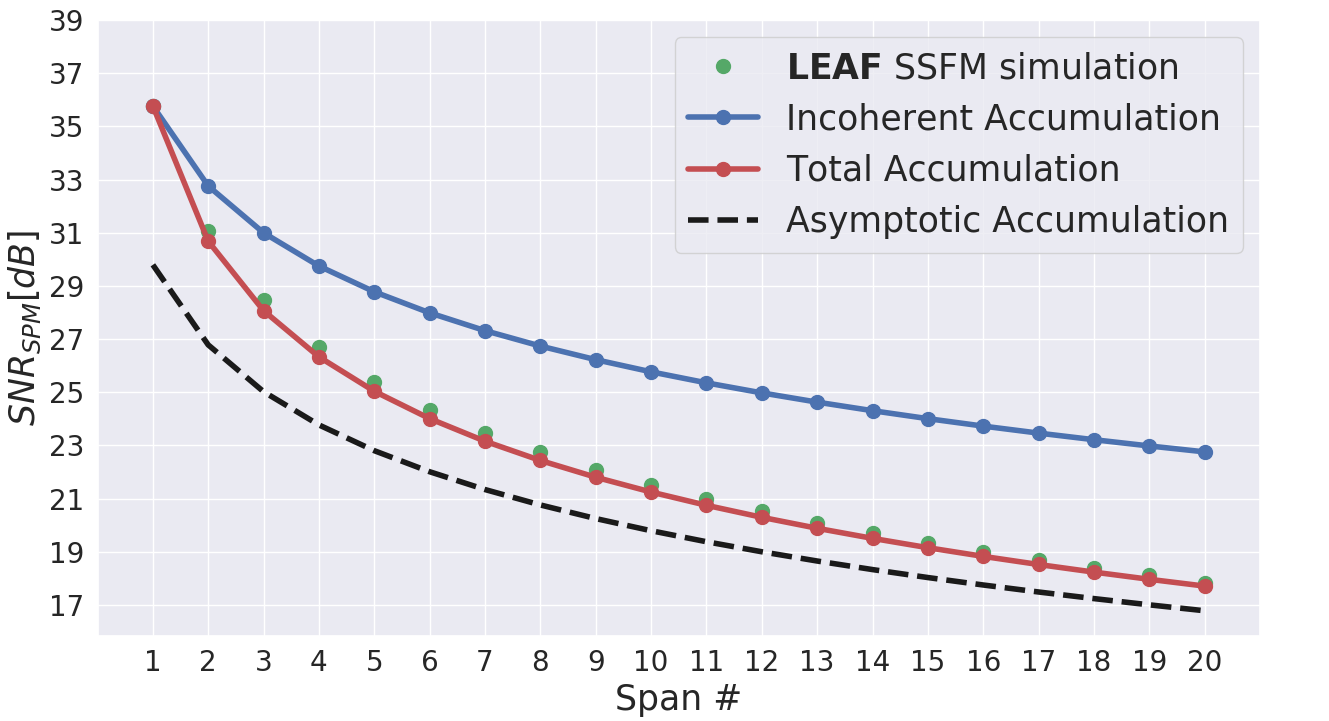}
   \end{subfigure}
   ~
   \centering
    \begin{subfigure}(c)
        \centering
        \includegraphics[width=0.45\textwidth]{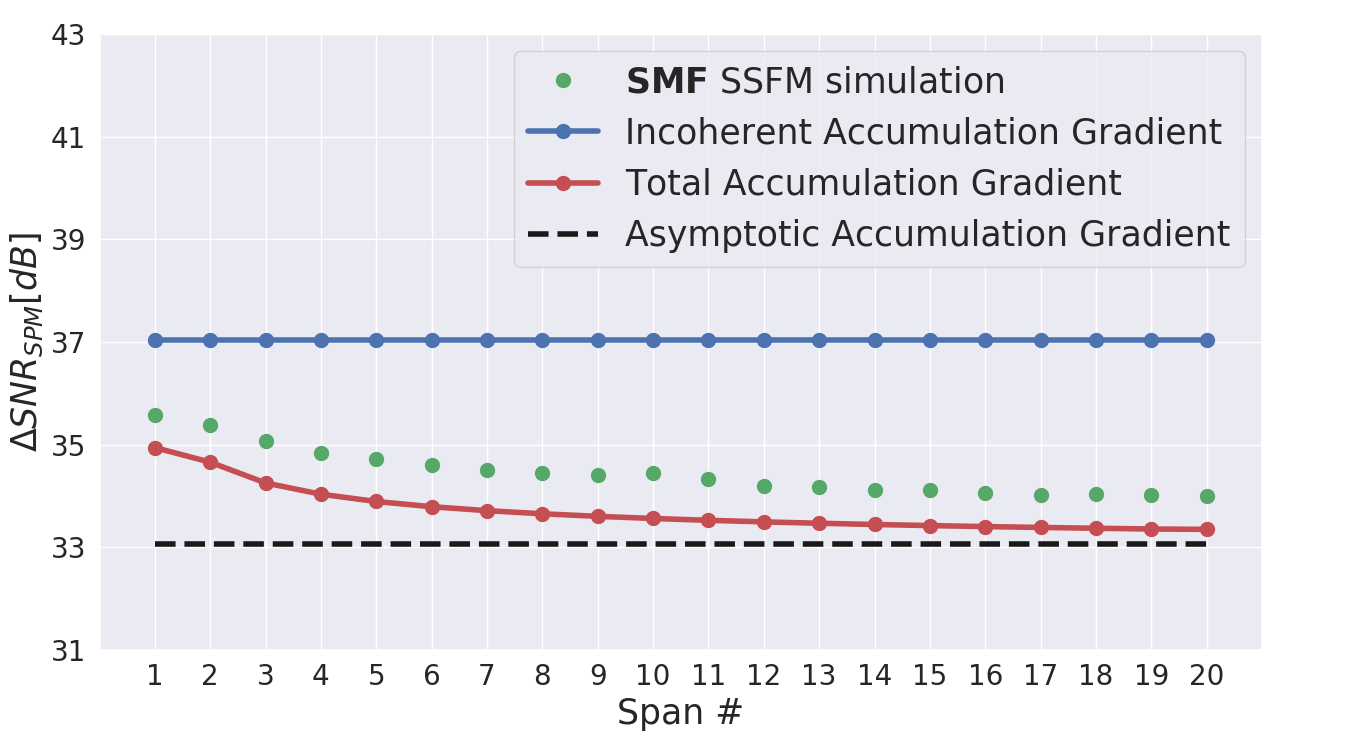}
    \end{subfigure}
    ~
    \begin{subfigure}(d)
        \centering
        \includegraphics[width=0.45\textwidth]{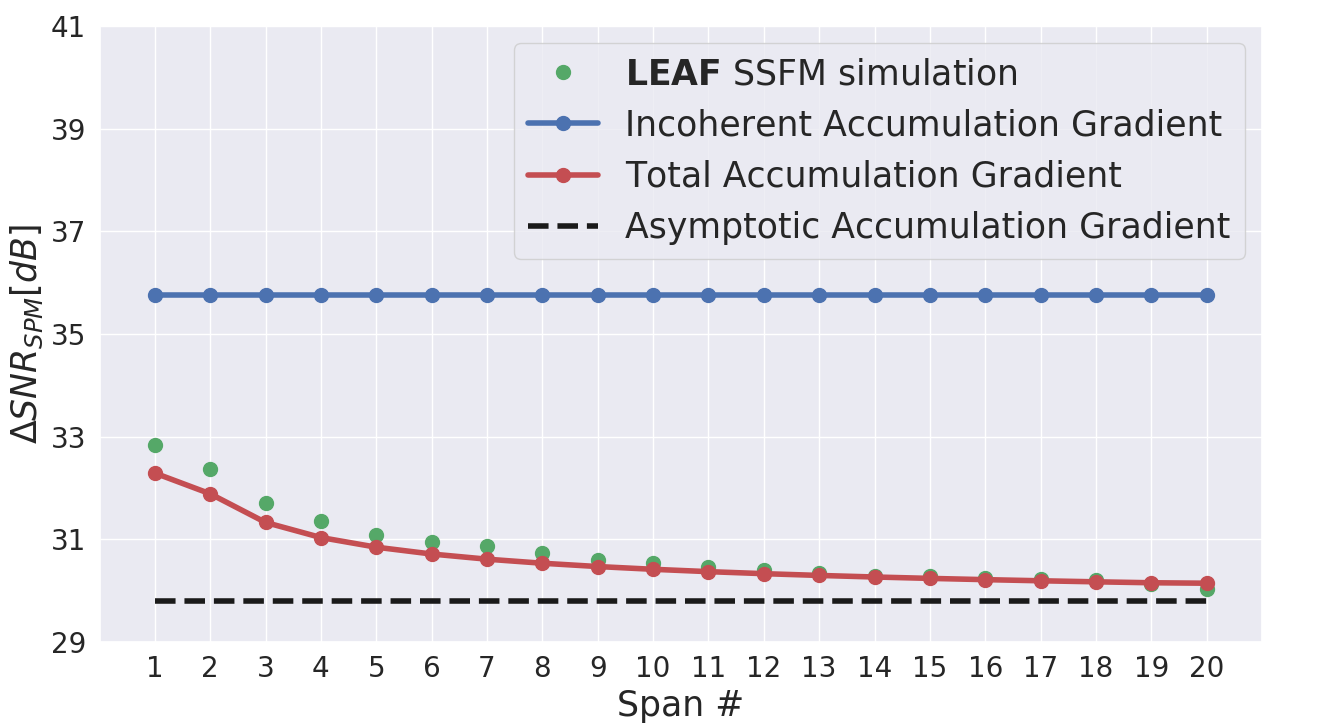}
   \end{subfigure}
   \caption{Simulated and predicted results for the SPM accumulation calculated for a $R_s=32$\,GHz channel, through 80\,km long spans in a periodical OLS for both SMF ($D=16.7$\,ps/nm$\cdot$km) and LEAF ($D=5.0$\,ps/nm$\cdot$km) fiber types. }
   \label{Fig:SPM Accumulation}
\end{figure*}
In a hybrid network scenario, data is transmitted with different spectral configurations and each OLS can be composed of various fiber types. These characteristics affect the generation of SPM and its accumulation in different ways. 
In our approach we analyze each span type within the network as an element of a periodic OLS. 
For this constraint, we investigate the asymptotic amount of SPM introduced by a single span and show how it can be exploited as a \emph{local} and conservative QoT estimation, in the disaggregated network abstraction.
The coherent accumulation is due to the intrinsic nature of the SPM. In fact, this quantity is generated by the NLI of the channel upon itself, therefore the SPM contributions at different spans are statistically correlated and sum coherently.
In the periodic scenario, the coherent accumulation term of the SPM power at the $n$th-span can be approximated by the following formula (see \cite{ucl14} and its bibliography for similar derivations):  
\begin{equation}
    \mathrm{COH}_{SPM}(n) \approx \mathrm{P}^{(n)}_{SPM}\:\: \frac{6}{5}\frac{2}{\sqrt{\theta}}\: \sum_{k=1}^{n-1}\frac{(n-k)}{k^{\frac{3}{2}}}\mathcal{C}(\sqrt{k\theta})=\mathrm{P}^{(n)}_{SPM}\:C^{(n)}\:,\quad \mathrm{with}\quad \mathcal{C}(\xi) = \int_{0}^{\xi} \mathrm{d}x \cos\left(\frac{\pi}{2} x^2\right)
    \label{Eq:Accumulation}
\end{equation}
where the dimensionless quantity $\theta$ defined in the previous section  groups all the physical parameters concerned in the accumulation and encodes the scale of the phenomenon. 

\begin{figure}[h!]
    \centering
    \includegraphics[width=0.75\textwidth]{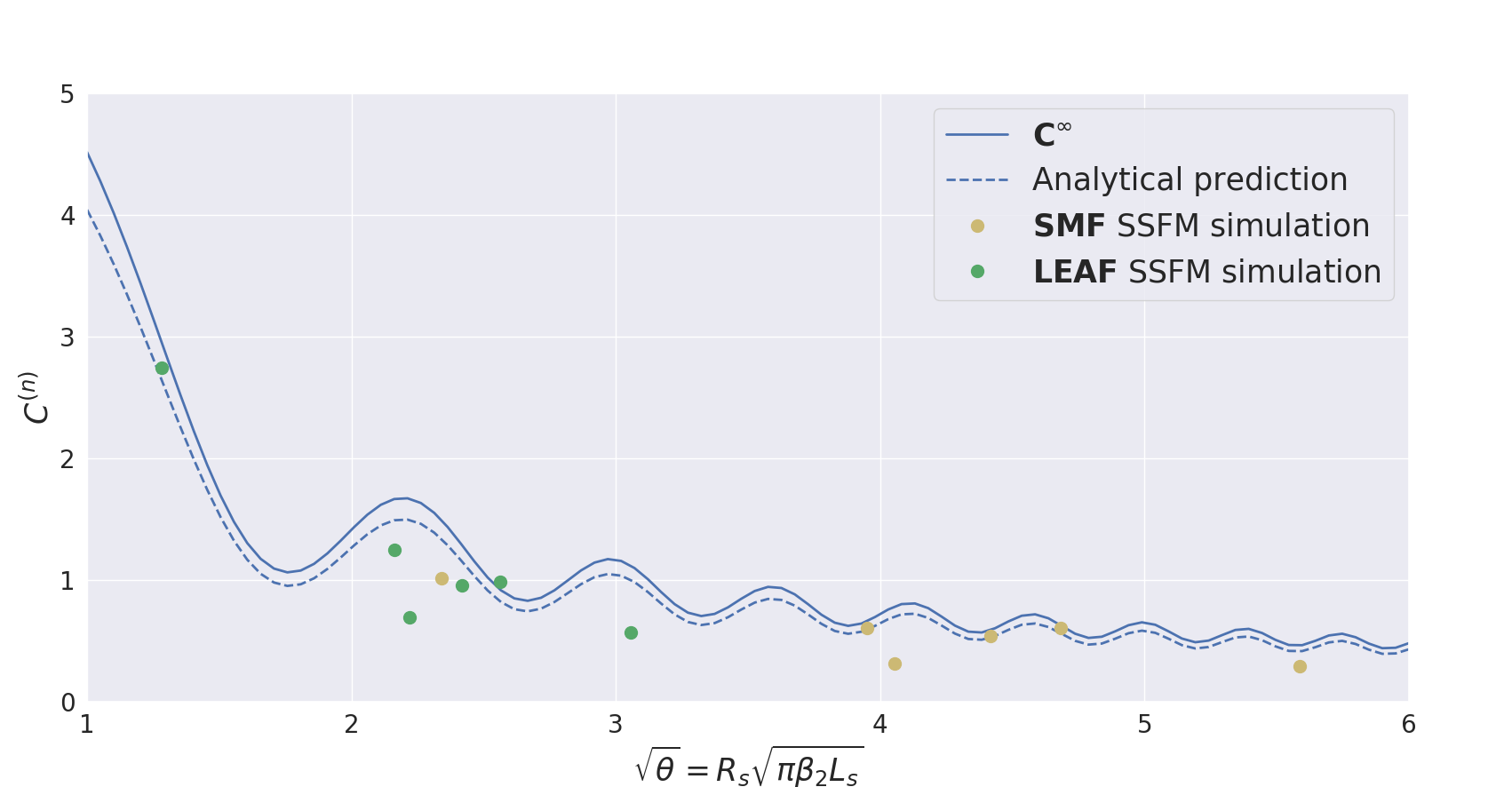}
    \caption{Simulation, analytical prediction and limit of the coherence coefficient $C^{(n)}$ along increasing $\theta$ values at $n = 20$.}
    \label{fig:Cinf}
\end{figure}

In order to validate the estimations produced by Eq.\ref{Eq:Accumulation}, we have performed several simulations using the Split Step Fourier Method (SSFM) with different combinations of lengths and symbol rate values for periodic OLS composed of both SMF an LEAF fibers. 
In Fig.\ref{Fig:SPM Accumulation}, we show a selection of two comparisons between the simulation and the analytical results, one for each fiber type. In both cases, we report the SPM accumulations by means of the respective SNR$_{SPM}$ (sub-figures (a) and (b)) and the variation of the differential amount of degradation, $\Delta$SNR$_{SPM}$, introduced after each span (sub-figures (c) and (d)).  
In particular, Fig.\ref{Fig:SPM Accumulation} shows that as the number of spans increases, so does the discrepancy between a model that considers only incoherent SPM accumulation and the simulated value, highlighting an underestimation of the QoT degradation rate. On the contrary, the evaluated values obtained via Eq.\ref{Eq:Accumulation} closely follow the simulation trends. In spite of the high accuracy, these analytical calculations need the information of the fiber parameters and the entire route history of the signal; these requirements avoid a blind, disjointed prediction of the QoT through a single span. Nevertheless, looking into the differential plots in Fig.\ref{Fig:SPM Accumulation}, it clearly arises that after a certain number of spans the coherent accumulation reaches its maximum and the new amount of SPM degradation introduced to each span remains constant. Therefore, to achieve a \emph{local} estimation, which has to be a conservative and accurate threshold of the SPM degradation, we propose the \emph{equivalent} model for the SPM generation obtained by means of a reasonable (high number of spans) limit of the $C^{(n)}$ coherence coefficient in Eq.\ref{Fig:SPM Accumulation}. In Fig.\ref{Fig:SPM Accumulation}, the effective results of this choice are shown (dashed black lines). Moreover, we have investigated the accuracy and reliability of this \emph{equivalent} disaggregated model for various values of $\theta$. In Fig.\ref{fig:Cinf}, we report the limit $C^{(\infty)}$, along incremental values of $\theta$, compared with both the simulation and the analytical evaluations. It is clear, at least in the investigated ranges of physical parameters, that this method permits a conservative estimation of the  SNR degradation due to SPM that does not diverge from the actual intensity of the phenomenon. 
\section{Conclusions}
In this study, we investigate the SPM disturbance generation through an OLS, focusing on its coherent accumulation. This impairment prevents a disaggregated abstraction of the spans as network elements, avoiding a reliable prediction of the signal degradation. To obtain a reliable and accurate QoT estimation of the SPM generation, we propose and validate an efficient approach based on an analytical derivation. 
The presented approach is suitable for a fully disaggregated network abstraction, as needed in planning, because it allows a standalone management of the SPM component of the NLI. Remarkably, this solution is conservative and completely independent of the other network elements.
\section*{Acknowledgment}
This work was partially funded by the EU H2020 within the ETN WON, grant agreement 814276.

\end{document}